\documentclass[final,3p,times,twocolumn]{elsarticle}
\biboptions{comma,sort&compress}
\usepackage{ecrc}
\usepackage{here}
\usepackage{graphicx}
\usepackage{epsfig}
\usepackage{epstopdf}
\usepackage{amsmath}
\usepackage{lipsum}

\def\nin{\noindent}
\def\beq{\begin{equation}}
\def\eeq{\end{equation}}
\def\bea{\begin{eqnarray}}
\def\eea{\end{eqnarray}}
\def\nnb{\nonumber}
\def\la{\langle}
\def\ra{\rangle}

\def\bma{\boldmath}															

\usepackage{graphicx}
\usepackage{here}
\def\beq{\begin{equation}}
\def\eeq{\end{equation}}
\def\bea{\begin{eqnarray}}
\def\eea{\end{eqnarray}}
\def\bq{\begin{quote}}
\def\eq{\end{quote}}
\def\ve{\vert}
\def\nnb{\nonumber}

\def\nnb{\nonumber}
\def\la{\langle}
\def\ra{\rangle}
\def\nin{\noindent}
\def\ba{\begin{array}}
\def\ea{\end{array}}

\def\als{\alpha_s}

\def\gg2{ \la\alpha_s G^2 \ra}
\def\gg3{g^3f_{abc}\la G^aG^bG^c \ra}
\def\ggg4{\la\als^2G^4\ra}

\def\beq{\begin{equation}}
\def\enq{\end{equation}}
\def\beqa{\begin{eqnarray}}
\def\enqa{\end{eqnarray}}
\def\nnb{\nonumber}

\def\MeV{\nobreak\,\mbox{MeV}}
\def\GeV{\nobreak\,\mbox{GeV}}
\def\keV{\nobreak\,\mbox{keV}}

\newcommand{\rag}{\rangle}
\newcommand{\lag}{\langle}

\def\gg{\lag g^{2}_{s} G^2 \rag}
\def\ggg{\lag g^{3}_{s}G^3\rag}

\volume{00}
\firstpage{1}
\journalname{Nuclear and Particle Physics Proceedings }
\runauth{D. Rabetiarivony}
\jnltitlelogo{Nuclear and Particle Physics Proceedings }

\begin{document}
\begin{frontmatter}

\title{Fully and Doubly-heavy four-quarks within QCD Laplace sum rule\tnoteref{text1}}

\author[label1]{R. M. Albuquerque}
\ead{raphael.albuquerque@uerj.br}
\address[label1]{Faculty of Technology,Rio de Janeiro State University (FAT,UERJ), Brazil}

\author[label2,label3]{S. Narison}
\ead{snarison@yahoo.fr}
\address[label2]{Laboratoire
Univers et Particules de Montpellier (LUPM), CNRS-IN2P3, Case 070, Place Eug\`ene
Bataillon, 34095 - Montpellier, France}
\address[label3]{Institute of High-Energy Physics of Madagascar (iHEPMAD), University of Ankatso, Antananarivo 101, Madagascar}

\author[label3]{A. Rabemananjara}
\ead{achrisrab@gmail.com} 

\author[label3]{D. Rabetiarivony\fnref{fn2}}
\fntext[fn2]{Speaker}
\ead{rd.bidds@gmail.com} 


\tnotetext[text1]{Talk given at QCD23 International Conference (10--14 July 2023, Montpellier--FR)}

\pagestyle{myheadings}
\markright{ }

\begin{abstract}
\noindent
We present a review of our results for the masses and couplings of the scalar fully heavy four-quarks and $T_{QQ\bar{q}\bar{q}'}\, (J^P=0^\pm , 1^\pm)$ tetraquarks states from (inverse) QCD Laplace sum rule (LSR), their ratios ${\cal R}$ and double ratio of sum rules (DRSR) within stability criteria and including Factorized Next-to-Leading Order (FNLO) Perturbative (PT) corrections. As the Operator Product Expansion (OPE) usually converges for $d\leqslant 6-8$, we evaluated the QCD spectral functions at Lowest Order (LO) of PT QCD and up to $\lag G^3 \rag$. Our results for the $0^{++}$ fully heavy four-quark states may explain the LHCb broad structure around $(6.2-6.7)\GeV$ which can be due to $\overline{\eta_c} \eta_c,~~ \overline{\chi_{c1}}\chi_{c1}$ and $\overline{J/\psi}J/\psi$ molecules or/and their analogue $S_c S_c,~~ A_cA_c$ and $V_cV_c$ tetraquarks. The peak at $(6.8-6.9)\GeV$ can be identified to the $\overline{\chi_{c0}}\chi_{c0}$ molecule or/and the $P_c P_c$ tetraquark state. Then, combining ${\cal R}$ and DRSR we focus on the analysis of the four-quark nature of $T_{cc\bar{q}\bar{q}'}$ $1^{\pm}$ and $0^{\pm}$ states. We show that combining ${\cal R}$ and DRSR can provide more precise results: $M_{T^{1^+}_{cc}}=3886(6)\MeV$ and $M_{T^{0^+}_{cc}}=3883(3)\MeV$. In a similar way, one obtain the improved mass predictions: $M_{T^{1^+}_{cc\bar{s}\bar{u}}}=3931(7)\MeV$, $M_{T^{0^+}_{cc\bar{s}\bar{u}}}=3983(7)\MeV$ and $M_{T^{0^+}_{cc\bar{s}\bar{s}}}=3993(11)\MeV$. From our estimates of the masses of the Pseudoscalar and Vector $T_{cc\bar{q}\bar{q}'}$ states, we observe that the interpolating currents lead to two classes: Class H (Heavy) states with masses around $6\GeV$ and Class L (Light) states around $(3.8-4.4)\GeV$ where the pseudoscalar (resp. all vector states) are below the $\overline{D}D_0,~~ \overline{D}_s D_{s0}$ (resp. $\overline{D}D_1,~~ \overline{D}_s D_{s1}$) open charm thresholds. Finally, we extend the whole study to the bottom sector and confront our results with the ones from different LSR predictions and some other approaches (lattice, potential models,$\cdots$) in the literature.
\end{abstract} 
\scriptsize
\end{frontmatter}
\section{Introduction}
In this talk, we summarize the results of our original works\,\cite{ANR3,ANR22,ANR23}.

\begin{figure}[hbt]
\begin{center}
\includegraphics[width=6.0cm]{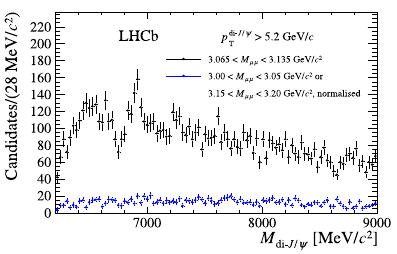}
\caption{\footnotesize Invariant mass spectrum of $J/\psi$-pair candidate. Data (resp. background regions) in black (resp. blue) signal.} 
\label{fig:lhcb4h}
\end{center}
\end{figure}

\begin{figure}[hbt]
\begin{center}
\includegraphics[width=6.0cm]{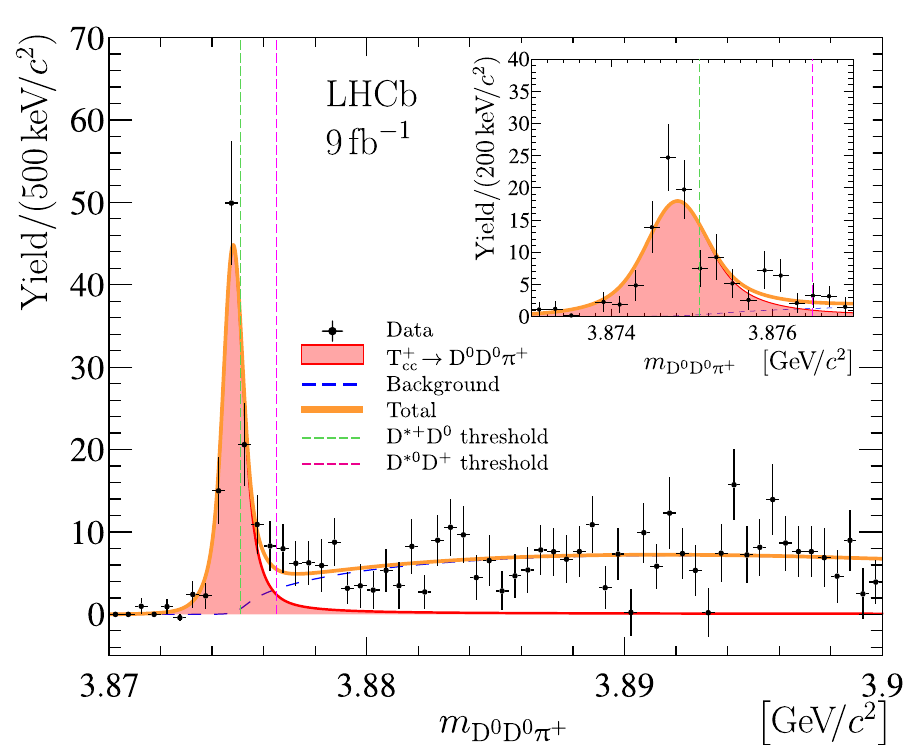}
\caption{\footnotesize Distribution of $D^0D^0\pi^+$ mass.} 
\label{fig:lhcbt}
\end{center}
\end{figure}

\nin Motivated by the LHCb-group discovery of exotic states around $(6.2-6.9)\GeV$ from the $J/\psi$-pair invariant mass spectrum\,\cite{LHCbQ} and at $3875\MeV$ from the $D^0D^0\pi^+$ mass distribution\,\cite{LHCbT} as we can see in Figs.\,\ref{fig:lhcb4h} and \ref{fig:lhcbt}, we use the inverse Laplace transform sum rules (LSR)\,\cite{BELL,BELLa,BNR,BERT,NEUF,SNR} and its ratios ${\cal R}$ to improve the existing QCD spectral sum rules (QSSR) \footnote{For reviews, see \cite{SVZa,Za,SNB1,SNB2,SNB3,CK,YND,PAS,RRY,IOFF,DOSCH,SNqcd22}} estimates of the masses and couplings of the scalar $(\bar{Q}Q)(\bar{Q}Q)$\footnote{$Q\equiv c,b$} molecules and $(\bar{Q}\bar{Q})(QQ)$ tetraquarks states. Then, to interpret the peak at $3875\MeV$ (Fig.\,\ref{fig:lhcbt}) which is a good candidate for $(I)J^P=(0)1^+$ state, we redo the analysis of the masses and couplings of $T_{QQ\bar q \bar q'}$ by combining the direct mass determinations from the ratios ${\cal R}$ with the ratio of masses from the double ratio of sum rules (DRSR).
\section{The Laplace sum rule}
\nin We shall work with the finite energy version of the QCD inverse Laplace sum rules and their ratios:
\bea
\hspace*{-0.2cm} {\cal L}^Q_n\vert_{\cal H}(\tau,\mu)&=&\int_{t^{2}_{0}}^{t_c} dt~t^n~e^{-t\tau}\frac{1}{\pi} \mbox{Im}~\Pi^{(1,0)}_{\cal H}(t,\mu)~;\nnb \\
 {\cal R}^c_{\cal H}(\tau)&=&\frac{{\cal L}^Q_{n+1}\vert_{\cal H}} {{\cal L}^Q_n\vert_{\cal H}},
\label{eq:lsr}
\eea
where $t_0=2M_Q+m_q+m_{q'}$ is the hadronic threshold, $M_Q$ ($Q\equiv c,b$) and $m_q,\,m_{q^{'}}$ ($q,q'\equiv u,d,s$) are respectively the heavy and light quarks masses (we shall neglect the $u,d$ quark masses), $\tau$ is the LSR variable, $n=0,\cdots$ is the degree of moments and $t_c$ is the "QCD continuum" threshold.\\
From the ratio of LSR, one can extract the mass squared at the optimization point $\tau_0$\,:
\beq
 {\cal R}^c_{\cal H}(\tau_0)= M_{\cal H}^2.
\eeq
We shall also work with the double ratio of sum rules (DRSR)\,\cite{DRSR88}\,:
\beq
r_{{\cal H'}/{\cal H}}(\tau_0)\equiv \sqrt{\frac{{\cal R}^c_{\cal H'}}{{\cal R}^c_{\cal H}}}=\frac{M_{\cal H'}}{M_{\cal H}},
\eeq
which can be free from systematics provided that ${\cal R}^c_{\cal H}$ and ${\cal R}^c_{\cal H'}$ optimize at the same values of $\tau$ and of $t_c$:
\beq
 \tau_0\vert_{\cal H}\simeq \tau_0\vert_{\cal H'}~,~~~~~~~~~~~~t_c\vert_{\cal H}\simeq t_c\vert_{\cal H'}~.
\eeq
The spectral function $\mbox{Im}~\Pi^{(1,0)}_{\cal H}(t,\mu)$ associated to the spin $1$ and $0$ interpolating operators can be evaluated from the two-point correlator:
\bea
\hspace*{-0.6cm}
\Pi^{(1,0)}_{\mathcal{H}}(q^2)\hspace*{-0.2cm}&=& \hspace*{-0.2cm} i \int \hspace*{-0.1cm} d^4 x\, e^{-i q x}\lag 0 \ve \mathcal{T} \mathcal{O}^{(1,0)}_{\mathcal{H}}(x)(\mathcal{O}^{(1,0)}_{\mathcal{H}}(x))^{\dag} \ve 0 \rag,\nnb \\
\eea
where $\mathcal{O}^{(1,0)}_{\mathcal{H}}(x)$ are the interpolating currents describing the four-quark states.

\nin The QCD expressions of the leading order (LO) spectral functions up to dimension 6 condensates, details about the FNLO PT corrections and the different QCD input parameters are given in Ref.\,\cite{SNqcd22,PICH,NPIV,HAGIWARA,RRY}.
\section{Optimization criteria}
In our previous works\,\cite{LNSR,ANRls,ANRTm,ANR21,ANR22p,ANR22pa,ANRR1,ANRR1a,ANRR2,ANR1,NR1,AFNR,SNX1,SNX2,SU3}, we have used the minimum sensitivity of the observables versus the variation of the external variables $\tau$, $t_c$ and $\mu$ to extract the optimal masses and couplings from the analysis.\\
\\ 
{\bf $\bullet$ $\tau$-stability:} The optimal result is obtained around a minimum or inflexion point. These optimal values of $\tau$ are equivalent to the so-called plateau used in the literature using the Borel $M^2\equiv 1/ \tau$ variable. At this $\tau$-stability region, one can check the lowest ground state dominance of the LSR and the convergence of the OPE.\\
{\bf $\bullet$ $t_c$-stability:} One expects it to be around the mass of the first radial excitation. To be conservative, we take $t_c$ from the beginning of $\tau$-stability until the $t_c$-stability starts to be reached. The $t_c$-stability region corresponds to a complete dominance of the lowest ground state in the QSSR.\\
{\bf $\bullet$ $\mu$-stability:} This is used to fix in a rigorous optimal way, the arbitrary subtraction constant appearing in the PT calculation of the Wilson coefficients and in the QCD input renormalized parameters. From our previous works, we have observed that the value of $\mu$-stability is (almost) universal for the different heavy-light four-quark states:
\beq
\mu_c=(4.65\pm 0.05)~{\rm GeV} ~~;~ \mu_b=(5.20\pm 0.05)~{\rm GeV}\, ,
\eeq
respectively for the charm and bottom channels.\\
{\bf $\bullet$} We shall also request that the Pole (P) contribution (resonance) to the spectral integral is larger than the QCD continuum (C) one:
\beq
R_{P/C}\equiv \frac{\int^{tc}_{t_0}dt e^{-t\tau}\mbox{Im}\Pi (t)}{\int^{\infty}_{t_c}dt e^{-t\tau}\mbox{Im}\Pi (t)}\geqslant 1.
\eeq
This condition is necessary to better restrict the conservative region of $(\tau,t_c)$. 
\section{The $(\bar{Q}Q)(\bar{Q}Q)$ molecules and $(\bar{Q}\bar{Q})(QQ)$ tetraquarks states}
We shall study the charm channels and their beauty analogue. As the shape of the different curves are similar, we shall illustrate it in the case of $\overline{\chi}_{c0}\chi_{c0}$. The final and conservative results are compiled in Table \ref{tab:fhresult}.
\subsection{Interpolating currents}
\nin We shall be concerned with the scalar interpolating currents.
\begin{table}[hbt]
\begin{center}
\setlength{\tabcolsep}{0.25pc}
{\scriptsize
\caption{\scriptsize Interpolating currents with a definite C-parity describing the molecules and tetraquarks states. $Q,q\equiv c,b$}
\begin{tabular}{ll}
&\\
\hline
\hline
Scalar States $(0^{++})$ & Currents $({\cal O}^{(0)}_{\cal H})$\\
 \hline\hline
Molecules & \\
$\overline{\chi}_{q0}\chi_{q0}$ & $(\bar{Q}Q)(\bar{Q}Q)$\\
$\overline{\eta}_q \eta_q$ & $(\bar{Q}\gamma_5 Q)(\bar{Q}\gamma_5 Q)$\\
$\overline{J/\psi}J/\psi$, $\overline{\upsilon}\upsilon$ & $(\bar{Q}\gamma_\mu Q)(\bar{Q}\gamma^\mu Q)$\\
$\overline{\chi}_{q1}\chi_{q1}$ & $(\bar{Q}\gamma_\mu\gamma_5 Q)(\bar{Q}\gamma^\mu\gamma_5 Q)$\\
Tetraquarks & \\
$\overline{P}_q P_q$ & $(Q^T_a C Q_b)(\bar{Q}_a C \bar{Q}^T_b)$\\
$\overline{s}_q S_q$ & $(Q^T_a C \gamma_5 Q_b)(\bar{Q}_a \gamma_5 C \bar{Q}^T_b)$\\
$\overline{A}_q A_q$ & $(Q^T_a C \gamma_\mu Q_b)(\bar{Q}_a \gamma^\mu C \bar{Q}^T_b)$\\
$\overline{V}_q V_q$ & $(Q^T_a C \gamma_\mu \gamma_5 Q_b)(\bar{Q}_a \gamma^\mu \gamma_5 C \bar{Q}^T_b)$\\
\hline\hline
\end{tabular}
\label{tab:fhcurrent}
}
\end{center}
\end{table}
\subsection{$\overline{\chi}_{c0}\chi_{c0}$ molecule state}
\begin{figure}[hbt]
\begin{center}
\includegraphics[width=6.0cm]{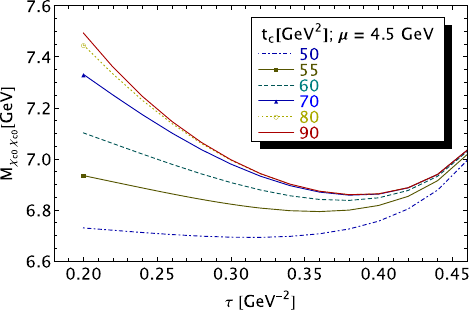}
\includegraphics[width=6.0cm]{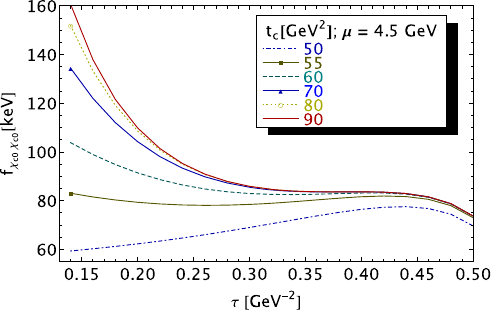}
\caption{\footnotesize The coupling $f_{\chi_{c0}\chi_{c0}}$ and mass $M_{\chi_{c0}\chi_{c0}}$ as a function of $\tau$ at NLO for \# values of $t_c$ and for $\mu$=4.5 GeV.} 
\label{fig:chic0}
\end{center}
\end{figure}

\nin We study the behavior of the coupling and mass in term of the LSR variable $\tau$ for different values of $t_c$ at NLO as shown in Fig.\,\ref{fig:chic0}. We consider as final results the mean of the value corresponding to the beginning of $\tau$ stability for $[t_c(\GeV),\tau (\GeV^{-2})]\simeq [55,\,0.36]$ and the one where the $t_c$ stability starts to be reached for $[t_c(\GeV),\tau (\GeV^{-2})]\simeq [70,\,0.38]$ from which we deduce:
\beq
M_{\chi_{c0}\chi_{c0}} = 6675(98)\MeV,~~f_{\chi_{c0}\chi_{c0}} = 69(21)\keV.
\eeq
\section{$T_{QQ\bar{q}\bar{q}'}$ states}
We extend the study to the case of the (pseudo)scalar and (axial-)vector $T_{QQ\bar{q}\bar{q}'}$ tetraquarks states using the ratios ${\cal R}$ of LSR and the double ratio DRSR. We do not consider the molecule assignment of the $T_{QQ\bar{q}\bar{q}'}$ states as the method cannot disentangle the molecule from the tetraquark states. As the analysis will be performed using the same techniques, we shall illustrate it in the case of $1^{+}$ $T_{cc\bar{u}\bar{d}}$. The final results are compiled in Tables\,\ref{tab:restccp}, \ref{tab:restccmc} and \ref{tab:restccmb}.
\subsection{Interpolating currents}
\nin We shall be concerned with the interpolating currents in Table\,\ref{tab:Tcurrent}. We have also re-analyzed the results of Ref.\,\cite{DCCZ} using the currents in Table\,\ref{tab:zhucurrent}.
\subsection{$T_{cc\bar{u}\bar{d}}$ or $T_{cc}(1^+)$ state from LSR at NLO}
\begin{figure}[hbt]
\begin{center}
\includegraphics[width=6.0cm]{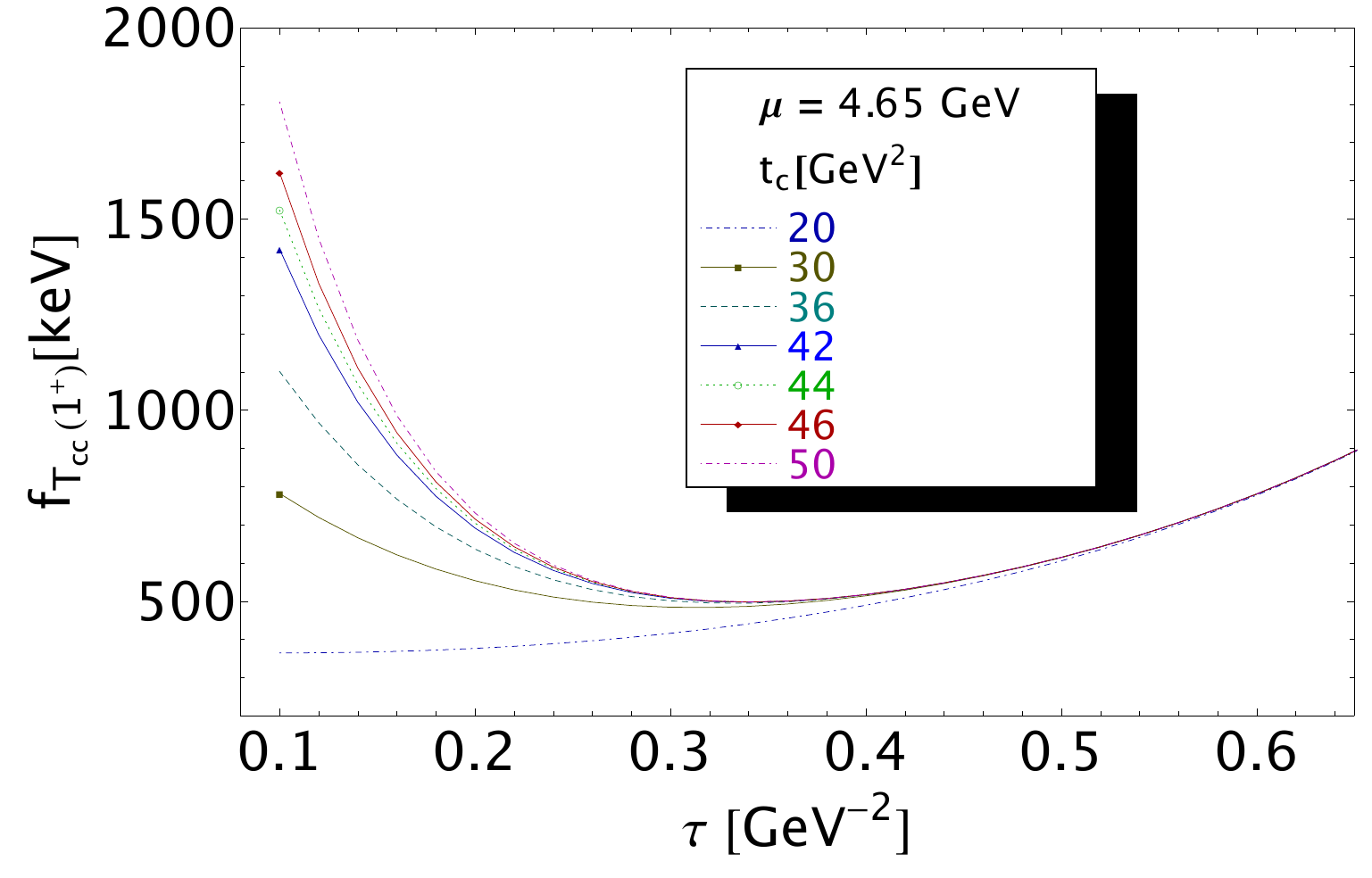}
\hspace*{0.5cm}
\includegraphics[width=6.0cm]{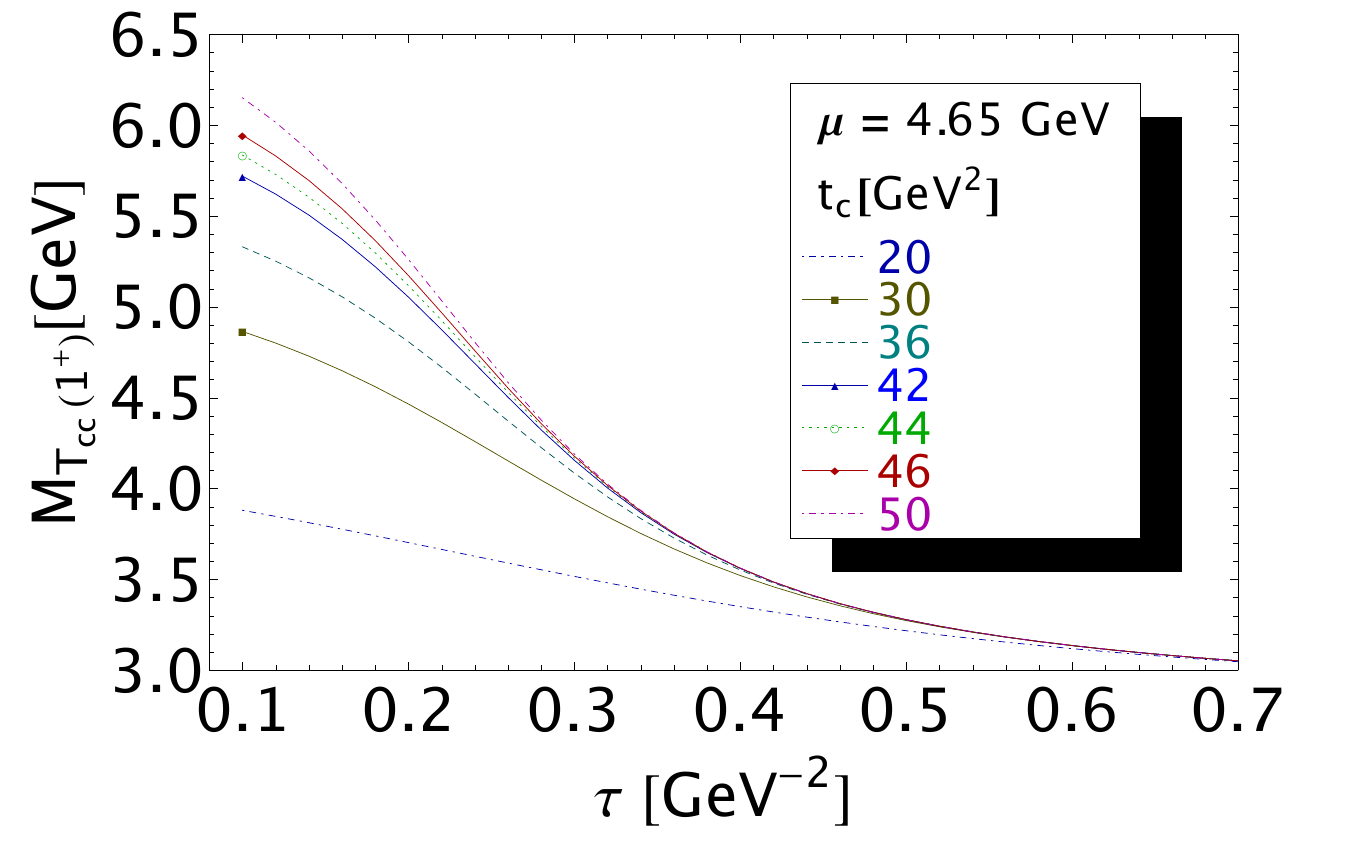}
\vspace*{-0.5cm}
\caption{\footnotesize  $f_{T_{cc}}(1^{+})$ and $M_{T_{cc}}(1^{+})$ as a function of $\tau$ at NLO for \# values of $t_c$ and for $\mu$=4.65 GeV.} 
\label{fig:tcc1p}
\end{center}
\vspace*{-0.5cm}
\end{figure}

The $\tau$ and $t_c$ behavior of the mass and coupling are shown in Fig.\,\ref{fig:tcc1p}. The optimal results are obtained for the set: $(\tau,t_c)=$(0.31,30) to (0.34,46) $(\rm GeV^{-2},\rm GeV^2)$ from which we deduce:
\beq
M_{T^{1^+}_{cc}} = 3885(123)\MeV,~~f_{T^{1^+}_{cc}} = 491(48)\keV .
\eeq
\subsection{$T_{cc}(1^+)$ state from DRSR}
\begin{figure}[hbt]
\begin{center}
\includegraphics[width=6.0cm]{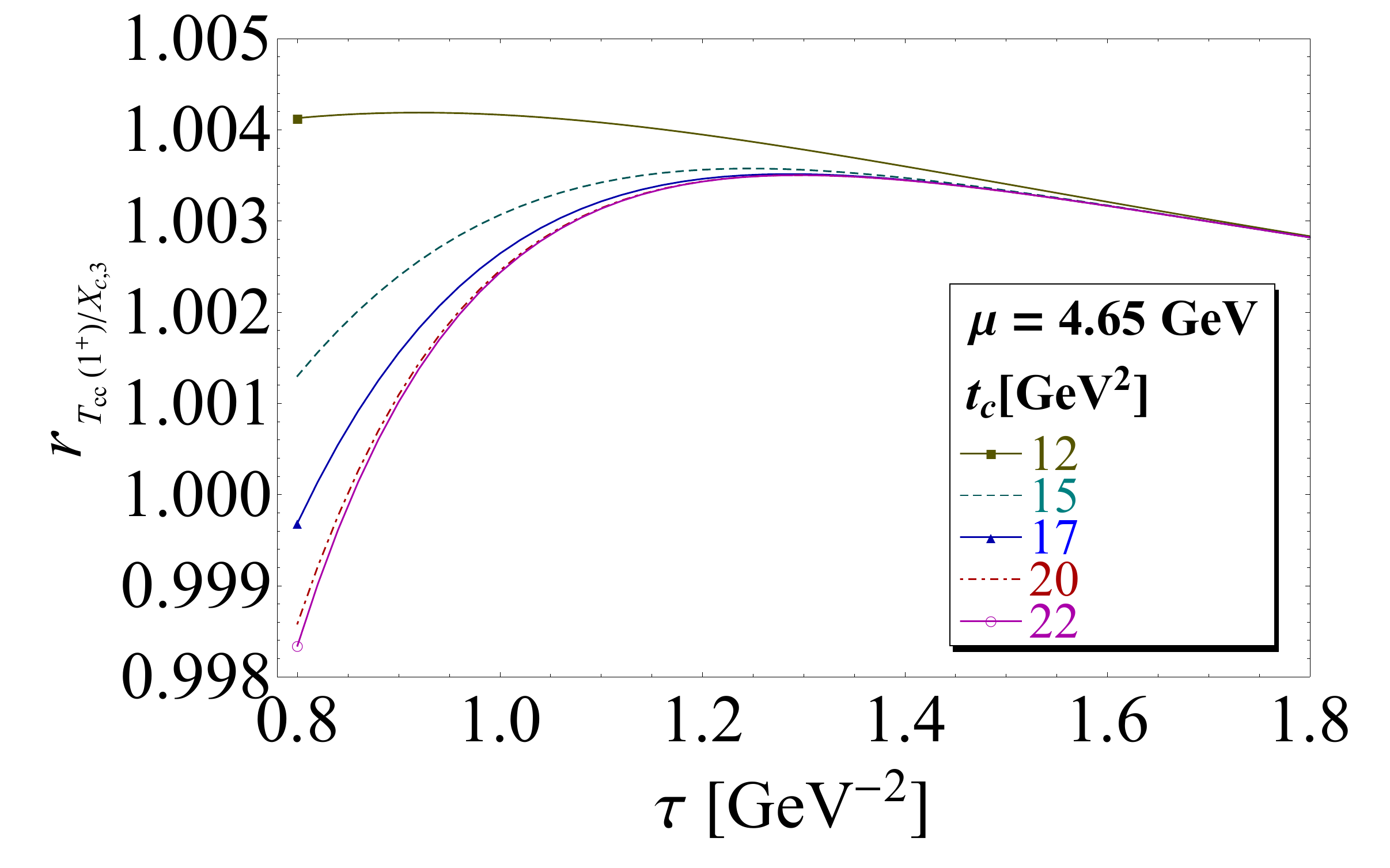}
\vspace*{-0.5cm}
\caption{\footnotesize Ratio of masses $r_{T^{1^+}_{cc}/X_c}$ as a function of $\tau$ at NLO for \# values of $t_c$ and for $\mu$=4.65 GeV.} 
\label{fig:tcc1pxc}
\end{center}
\vspace*{-0.5cm}
\end{figure}

The optimal result is obtained for the set: $(\tau,t_c)=$(0.31,30) to (0.34,46) $(\rm GeV^{-2},\rm GeV^2)$ (Fig.\,\ref{fig:tcc1pxc}) from which we deduce:
\beq
r_{T^{1^+}_{cc}/X_c}=1.0035(10) \Rightarrow M_{T^{1^+}_{cc}} = 3886(4)\MeV,
\eeq
where we have used the experimental mass of the $X_c(3872)$\,\cite{X3872-EXP}.
\subsection*{$\lozenge$ Final value of $M_{T^{1^+}_{cc}}$ from LSR$\oplus$DRSR}
\nin
As a final value of $M_{T^{1^+}_{cc}}$, we take the mean of the previous two determinations:
\beq
M_{T^{1^+}_{cc}}=3886(4)~\rm{MeV}.
\eeq
This estimation is comparable with the LHCb data $T_{cc}(1^+)= 3875\MeV$ which is slightly above the $D^*D$ threshold of $3877\MeV$. 
\section{Summary}
We have presented improved predictions of QSSR for the masses and couplings of $J^{PC}=0^{++}$ fully heavy molecules and tetraquarks states using LSR at NLO of PT series and including non-perturbative contribution condensates up to dimension $d\leqslant 6$. We have also improved the existing QSSR results of (pseudo)scalar and (axial-)vector $T_{QQ\bar{q}\bar{q}'}$ states by combining the direct mass determinations from the ratios ${\cal R}$ of LSR with the ratio of masses from DRSR. Our analysis has been done within stability criteria with respect to the LSR variable $\tau$, the QCD continuum threshold $t_c$ and the subtraction constant $\mu$. To better restrict the stability region of $(\tau,t_c)$, we have also requested that the Pole contribution is larger than the QCD continuum $(R_{P/C}\geqslant 1)$.\\
-- From the previous analysis, we can conclude that the broad structure around $(6.2\sim 6.7)\GeV$ might be explained by the $\overline{\eta_c} \eta_c,~~ \overline{\chi_{c1}}\chi_{c1}$ and $\overline{J/\psi}J/\psi$ molecules or/and their analogue $S_c S_c,~~ A_cA_c$ and $V_cV_c$ tetraquarks. The narrow structure at $6.9\GeV$ can be identified with a $\overline{\chi_{c0}}\chi_{c0}$ molecule or $P_c P_c$ tetraquark.\\
-- The $\overline{\chi_{c1}}\chi_{c1}$ mass is below the $\chi_{c1}\chi_{c1}$ threshold, while all of the predicted masses of the beauty state are above the $\eta_b\eta_b$ and $\Upsilon(1S)\Upsilon(1S)$ thresholds.\\
-- Our predictions for the $X_c$ and $Z_c$ states can reproduce quite well the observed masses of the $X_c(3872)$ and $Z_c(3900)$. However, as mentioned above and in\,\cite{ANR22}, our method cannot distinguish the molecule and four-quark nature of a state. At the present stage, we can only provide a description of these states as \emph{Tetramole} $(\cal T)$.\\
-- Comparing our results with different QSSR predictions in the literature, one can conclude that within the errors, there are (almost) good agreement among different determinations. Most of the approaches predict the scalar and axial $T_{ccqq'}$ states to be above the hadronic thresholds while the $T_{bbqq'}$ is expected to be below the physical thresholds.\\
-- Confronted to different approaches in the literature (lattice calculations\,\cite{JUN,MALT,MOHAN,LESK},  light front holographic\,\cite{DOSCH2}, quark and potential models $\oplus$heavy quark symmetry\,\cite{MENG,ROSNER,QUIGG,BRAATEN,CHENG,RICHARD2,ZHUMODEL,WU,BICUDO}) one can also notice that our LSR$\oplus$DRSR predictions are (almost) in agreement (within the errors). The $0^+$ and $1^+$ $T_{cc}$ states is expected to be above the physical threshold while the $T_{bb}$ ones are grouped around the threshold.\\
-- Our different predictions for $0^+$ and $1^+$ states from LSR$\oplus$DRSR are clustered in the range $-250$ to $+150$ MeV of the physical thresholds.\\
-- From our results for the masses, we can divide the pseudoscalar and vector states into two classes: The Heavy (Class H) and Light (Class L) states. The masses of the Class H $T_{cc\bar{q}\bar{q}'}$ (resp. $T_{bb\bar{q}\bar{q}'}$) states are around 6 (resp. 13)$\GeV$ which are about $2\GeV$ higher than the ones of their axial and scalar partners. This findings goes in line with our previous predictions for XYZ states\,\cite{SNX1,SNX2,SU3,ANR21}. For the Class L states the masses of the charm channels are around $(3.8\sim 4.4)\GeV$, while for the bottom ones they are about $10.4\GeV$ and all of them are below the open beauty thresholds.\\

Due to space limitation, we did not include in this paper some results like the first radial excitations ones,$\cdots$ For more details, refer to the original works\,\cite{ANR3,ANR22,ANR23}.

\begin{table*}[hbt]
\begin{center}
\setlength{\tabcolsep}{2.9pc}
\caption{\scriptsize Interpolating operators describing the $ZXT$ states where $b=0$ is the optimized mixing parameter\,\cite{ANR21}.}
{\scriptsize
  \begin{tabular}{lll}
&\\
\hline
\hline
States&$I(J^P)$& $\bar 3_c3_c$ Four-quark  Currents\\
\hline
\\
$Z_c$ &$(1^{+})$
&$ {\cal O}_{A_{cq}}=\epsilon_{ijk}\epsilon_{mnk}\big{
[}(q^T_i\,C\gamma_5\,c_j)(\bar q'_m \gamma_\mu C\, \bar c_n^T)
\,+b,(q^T_i\,C\,c_j)(\bar q'_m\gamma_\mu \gamma_5 C\, 
\bar c_n^T)\big{]}$\\
&&$ {\cal O}_{D^*_qD_q}=(\bar c\gamma_\mu q)(\bar q'\,i\gamma_5c)$  \\
\\
 $X_c $&$(1^{+})$
&$ {\cal O}^3_{X} = \epsilon_{i j k} \:\epsilon_{m n k} \big{[}\left(
  q_i^T\, C \gamma_5 \,c_j \right) \left( \bar{c}_m\, \gamma^\mu
  C \,\bar{q}_n^T\right) + \left(
  q_i^T\, C \gamma^\mu \,c_j \right) \left( \bar{c}_m\, \gamma_5
  C \,\bar{q}_n^T\right)\big{]}$ \\
 && $ {\cal O}^6_{X} = \epsilon_{i j k} \:\epsilon_{m n k}\big{[} \left(
  q_i^T\, C \gamma_5 \lambda_{ij}^a\,c_j \right) \left( \bar{c}_m\, \gamma^\mu
  C\lambda_{mn}^a \,\bar{q}_n^T\right) +  \left(
  q_i^T\, C \gamma^\mu \lambda_{ij}^a\,c_j \right) \left( \bar{c}_m\, \gamma_5
  C \lambda_{mn}^a\,\bar{q}_n^T\right)\big{]}$ \\
  && $ {\cal O}_{D^*_qD_q}= \frac{1}{\sqrt{2}}\big{[}(\bar q\gamma_5 c) (\bar c\gamma_\mu q) -  (\bar q\gamma_\mu c) (\bar c\gamma_5 q)\big{]} $ \\
 & & $ {\cal O}_{\psi\pi} = (\bar c\gamma_\mu\lambda^a c)(\bar q\gamma_5\lambda^a q)$\\
  \\

 $T_{cc\bar u\bar d}$&$0(1^{+})$
 &  $ {\cal O}_T^{1^+} = \frac{1}{\sqrt{2}}\epsilon_{i j k} \:\epsilon_{m n k} \left(
 c_i^T\, C \gamma^\mu \,c_j \right) \big{[} \left( \bar{u}_m\, \gamma_5
  C \,\bar{d}_n^T\right) -  \left( \bar{d}_m\, \gamma_5
  C \,\bar{u}_n^T\right)\big{]}$ \\
$T_{cc\bar u\bar s}$&$\frac{1}{2}(1^{+})$
 &  $ {\cal O}_{T^{1^+}_{us}}   = \epsilon_{i j k} \:\epsilon_{m n k}
    \left( c_i \, C \gamma^{\mu } c_j^T \right) 
    \left( \bar{u}_m \,\gamma_5 C \bar{s}_n^T \right)$\\
  
  \\
$T_{cc\bar u\bar d}$&$1(0^{+})$
 &  $ {\cal O}_T^{0^+} = \frac{1}{\sqrt{2}}\epsilon_{i j k} \:\epsilon_{m n k} \left(
 c_i^T\, C \gamma^\mu \,c_j \right) \big{[} \left( \bar{u}_m\, \gamma_\mu
  C \,\bar{d}_n^T\right) + \left( \bar{d}_m\, \gamma_\mu
  C \,\bar{u}_n^T\right)\big{]}$\\
  
    $T_{cc\bar u\bar s}$&$\frac{1}{2}(0^{+})$
 &  $ {\cal O}_{T^{0^+}_{us}}   = \epsilon_{i j k} \:\epsilon_{m n k}
    \left( c_i \, C \gamma_{\mu } c_j^T \right) 
    \left( \bar{u}_m \,\gamma^{\mu } C \bar{s}_n^T \right)$\\

  $T_{cc\bar s\bar s}$&$0(0^{+})$
 &  $ {\cal O}_T^{0^+}    = \epsilon_{i j k} \:\epsilon_{m n k}
    \left( c_i \, C \gamma_{\mu } c_j^T \right) 
    \left( \bar{s}_m \,\gamma^{\mu } C \bar{s}_n^T \right)$\\
\\
    $T_{cc\bar u\bar d}$& $(0^-)$
 &  $ {\cal O}_{T_{ud}^{0^-} }= \frac{1}{\sqrt{2}}\epsilon_{i j k} \:\epsilon_{m n k} \left(
 c_i^T\, C  \gamma^\mu \,c_j \right) \big{[} \left( \bar{u}_m\,\gamma_5\, \gamma_\mu
  C \,\bar{d}_n^T\right) - \left( \bar{d}_m\,\gamma_5\, \gamma_\mu
  C \,\bar{u}_n^T\right)\big{]}$\\
  
    $T_{cc\bar u\bar s}$& $(0^-)$
 &  $ {\cal O}_{T^{0^-}_{us}}   = \epsilon_{i j k} \:\epsilon_{m n k}
    \left( c_i \, C \gamma_{\mu } c_j^T \right) 
    \left( \bar{u}_m\,\gamma_5 \,\gamma^{\mu } C \bar{s}_n^T \right)$\\
   \\
    $T_{cc\bar u\bar d}$&$(1^{-})$
 &  $ {\cal O}_{T_{ud}^{1^-}} = \frac{1}{\sqrt{2}}\epsilon_{i j k} \:\epsilon_{m n k} \left(
 c_i^T\, C \gamma^\mu \,c_j \right) \big{[} \left( \bar{u}_m\, 
  C \,\bar{d}_n^T\right) -  \left( \bar{d}_m\, 
  C \,\bar{u}_n^T\right)\big{]}$\\
      $T_{cc\bar u\bar s}$&$(1^{-})$
 &  $ {\cal O}_{T^{1^-}_{us}}   = \epsilon_{i j k} \:\epsilon_{m n k}
    \left( c_i \, C \gamma^{\mu } c_j^T \right) 
    \left( \bar{u}_m \, C \bar{s}_n^T \right)$\\
   \\
   \hline\hline
\end{tabular}
}
\label{tab:Tcurrent}
\end{center}
\end{table*}

\begin{table*}[hbt]
\begin{center}
\setlength{\tabcolsep}{3.2pc}
 \caption{\scriptsize Interpolating operators describing the $T_{cc\bar q\bar q'}$-like states used in Ref.\,\cite{DCCZ}.}
    {\scriptsize
  \begin{tabular}{lll}
&\\
\hline
\hline
States&  Four-quark  Currents & Light quarks $q$\\
\hline
{\it  Pseudoscalar $0^-$}\\
 $T_{cc\bar q\bar q}$
 &  $ \eta_1 =  \left(
 c_a^T\, C \,c_b \right) \big{[} \left( \bar{q}_a\gamma_5 \, 
  C \,\bar{q}_b^T\right) +  \left( \bar{q}_b\gamma_5\, 
  C \,\bar{q}_a^T\right)\big{]}$&$u,d,s$\\

  $T_{cc\bar q\bar q}$
 &  $\eta_2 = \left(
 c_a^T\, C  \gamma_5 \,c_b \right) \big{[} \left( \bar{q}_a\,
  C \,\bar{q}_b^T\right) + \left( \bar{q}_b\,
  C \,\bar{q}_a^T\right)\big{]}$&$u,d,s$\\

 $T_{cc\bar q\bar u}$
 &  $\eta_4= \left(
 c_a^T\, C  \gamma_\mu \,c_b \right) \big{[} \left( \bar{u}_a\,\gamma^\mu\gamma_5\, 
  C \,\bar{q}_b^T\right) - \left( \bar{u}_b\,\gamma^\mu\gamma_5\, 
  C \,\bar{q}_a^T\right)\big{]}$&$d,s$\\
  $T_{cc\bar q\bar u}$
 &  $\eta_5 = \left(
 c_a^T\, C  \gamma_\mu \gamma_5\,c_b \right) \big{[} \left( \bar{u}_a\,\gamma^\mu\, 
  C \,\bar{q}_b^T\right) + \left( \bar{u}_b\,\gamma^\mu\, 
  C \,\bar{q}_a^T\right)\big{]}$&$d,s$\\
\hline
{\it Vector $1^-$}\\
 $T_{cc\bar q\bar q}$
 &  $ \eta_1 =  \left(
 c_a^T\, C \gamma^\mu\gamma_5 \,c_b \right) \big{[} \left( \bar{q}_a\gamma_5 \, 
  C \,\bar{q}_b^T\right) +  \left( \bar{q}_b\gamma_5\, 
  C \,\bar{q}_a^T\right)\big{]}$&$ u,d,s$ \\
   
$T_{cc\bar q\bar q}$
 &  $\eta_2 = \left(
 c_a^T\, C  \gamma_5 \,c_b \right) \big{[} \left( \bar{q}_a\,\gamma_\mu\gamma_5\, 
  C \,\bar{q}_b^T\right) + \left( \bar{q}_b\,\gamma_\mu\gamma_5\, 
  C \,\bar{q}_a^T\right)\big{]}$&$u,d,s$\\
  
   $T_{cc\bar u\bar q}$
 &  $ \eta_5 =  \left(
 c_a^T\, C \gamma^\mu \,c_b \right) \big{[} \left( \bar{u}_a \, 
  C \,\bar{q}_b^T\right) -  \left( \bar{u}_b\, 
  C \,\bar{q}_a^T\right)\big{]}$&$ d,s$\\
 
    $T_{cc\bar u\bar q}$
 &  $ \eta_6 =  \left(
 c_a^T\, C \,c_b \right) \big{[} \left( \bar{u}_a \, 
  \gamma^\mu C \,\bar{q}_b^T\right) +  \left( \bar{u}_b\, 
  \gamma^\mu C \,\bar{q}_a^T\right)\big{]}$&$d,s$\\
   \hline\hline
\end{tabular}}
\label{tab:zhucurrent}
\end{center} 
\end{table*}

\begin{table*}[hbt]
\begin{center}
\caption{\scriptsize  Scalar fully heavy molecules/tetraquarks couplings and masses predictions from LSR at NLO. $\vert \Delta \mu \vert=0.20$ (resp. $0.25$) for the charm (resp. beauty) case. $\vert \Delta \tau \vert=0.02\GeV^{-2}$. In the case of asymmetric errors, we take the mean value.}

\includegraphics[width=16.5cm]{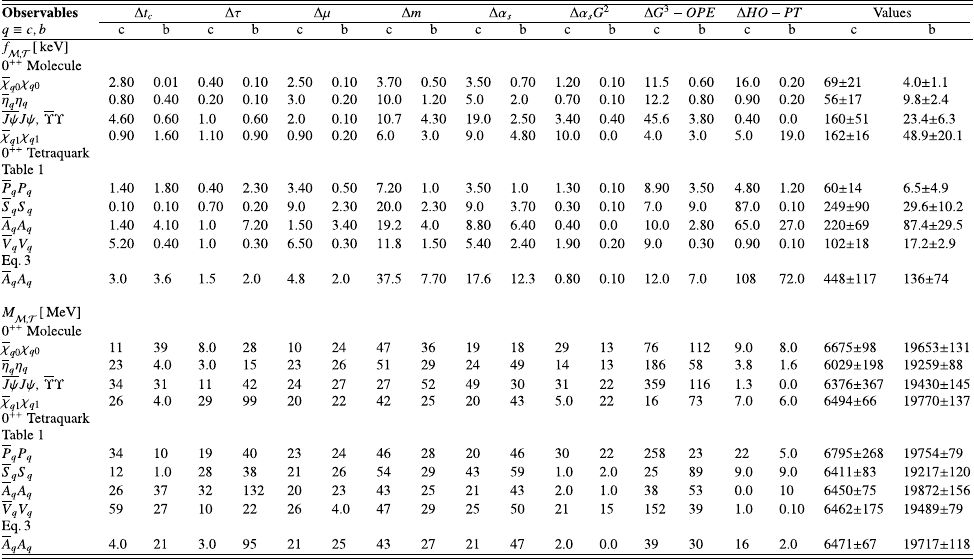}
\label{tab:fhresult}
\end{center}
\end{table*}

\begin{table*}[hbt]
\begin{center}
\setlength{\tabcolsep}{1.15pc}
\caption{\scriptsize Summary of the results of the $XZT$ states masses in units of MeV obtained from LSR and DRSR. Our final values are in the column ``LSR $\oplus$ DRSR" which are the mean from LSR with the ones deduced from DRSR.}
\vspace*{0.5cm}
{\scriptsize
\begin{tabular}{llllll  ll   l }
\hline\hline

&&&&\multicolumn{4}{c}{Mass}& \multicolumn{1}{l}{$\Delta E_B$}\\
\cline{5-9} 
States&$J^{P}$& 
Decay&Thresholds&\multicolumn{1}{l}{Data}&\multicolumn{1}{l}{Config.}&\multicolumn{1}{l}{LSR}
&\multicolumn{2}{c}{LSR $\oplus$ DRSR}\\
 \hline
     $Z_c$&$1^{+}$&$\bar{D^0}D^{*+}$& 3876&3900&$D^*D$&3912(61)\, \cite{ANR21}&&\\
     &&&&& $\bar  3_c3_c$&3889(58)\cite{ANR21}\\
         &&&&& ${\cal T}_{Z_c}$&&\it 3900(42)\,\cite{ANR21}&+24(42)\\
  $Z_b$    &$1^{+}$&$\bar B^0B^{*+}$&10605&& ${\cal T}_{Z_b}$&&\it 10579(99)\,\cite{ANR21}&--26(99)\\    
  \\
  $X_c$&$1^{+}$&$\bar{D^0}D^{*+}$& 3876&3872&$\bar  3_c3_c$&3876(76)& \\
    &&&&& $\bar  6_c6_c$&&3864(76)&\\
    &&&&& $ {\psi\pi}$&&3889(76)&\\
     &&&&& $ {D^*D}$&&3912(61)&\\
    &&&&& ${\cal T}_{X_c}$&&\it 3876(44)&+0(44)\\
     $X_b$&$1^{+}$&$\bar B^0B^{*+}$&10605&&$\bar  3_c3_c$&10545(131)&&--60(131) \\
\\
  $T_{cc\bar u\bar d}$&$1^{+}$&$\bar{D^0}D^{*+}$& 3876&3875&$\bar  3_c3_c$&3885(74)&3886(4)&+14(6)\\
  $T_{cc\bar u\bar s}$&$1^{+}$&$\bar {D^0_s}D^*$& 3975&&--
&3940(89)&3931(7)& -44(7)\\

$T_{cc\bar q\bar q}$&$0^{+}$&$\bar {D^0}D^0$& 3730&&--
&3882(81)&3883(3)& +153(3)\\
  $T_{cc\bar u\bar s}$&$0^{+}$&$\bar {D^0_s}D$& 3833&&--
&3936(90)&3927(6)&+94(6)\\

$T_{cc\bar s\bar s}$&$0^{+}$& $ {D^+_s}D^-_s$& 3937&&--&4063(72)&3993(11)&+56(11)\\
\\
$T_{bb\bar u\bar d}$&$1^{+}$& $\bar{B^0}B^{*+}$& 10605&&--&10441(147)&10501(98)& -104(98)\\
$T_{bb\bar u\bar s}$&$1^{+}$& $\bar{B^0_s}B^{*+}$& 10692&&--&10476(154)&10521(83)& -171(83)\\

$T_{bb\bar q\bar q}$&$0^{+}$& $\bar {B^0}B^0$ &10559&&--&10419(146)&10484(63)& -75(63)\\
$T_{bb\bar d\bar s}$&$0^{+}$& $\bar {B^0_s}B^0$ &10646&&--&10454(153)&10511(58)& -135(56)\\
$T_{bb\bar s\bar s}$&$0^{+}$& $ {\bar B^0_s}B^0_s$& 10734&&--&10538(129)&10567(57)& -167(57)\\

\hline\hline
\end{tabular}
} 
\label{tab:restccp}
\end{center}
\end{table*}
\begin{table*}[hbt]
\begin{center}
\setlength{\tabcolsep}{0.8pc}
\caption{ \scriptsize $T_{cc\bar{q}\bar{q}'}$ Predictions of the couplings [keV] and masses [MeV] compared with the ones in Ref.\,\cite{DCCZ}.}  
{\scriptsize
  \begin{tabular}{lllllll | lll}
&\\
\hline
\hline
\multicolumn{7}{c}{\bf Our Work}&\multicolumn{3}{c}{Ref.\,\cite{DCCZ}}\\
\hline
States&Current  &$t_c$ [GeV$^2$] &$\tau$  [GeV$^{-2}$]  & $f^{NLO} _{T_{ccqq'}}$&$\tau$  [GeV$^{-2}$] &$M^{NLO}_{T_{ccqq'}}$&$t_c$ & $\tau$  [GeV$^{-2}$] &$M^{LO}_{T_{ccqq'}}$\\ 
 \hline 
 \bf Class H &&&&&&&\\
{ \bma  $0^-$} &&&&&&&\\
$T_{cc\bar u\bar d}$&$ {\cal O}_T^{0^-},\eta_4$&$50 \to 60$&$0.12\to 0.18$&1559(133)&$0.16\to 0.18$&6303(123)&22&$0.29\to 0.37$& 4430(130) \\
$T_{cc\bar d\bar d}$ & $\eta_2$&$48\to 60$&$0.14\to 0.18$&1348(165)&$0.16\to 0.18$&6267(148)&&&-- \\
$T_{cc\bar u\bar s}$& $ {\cal O}_{T^{0^-}_{us}},\eta_4$&$44\to 58$&$0.16\to 0.20$&983(147)&$0.15\to 0.20$&6102(172)&&&-- \\
$T_{cc\bar s\bar s}$ & $\eta_2$&$42\to 58$&$0.16\to 0.22$&1164(168)&$0.18\to 0.22$&5946(180)&&&-- \\
{ \bma $1^-$} &&&&&&&\\
 $T_{cc\bar u\bar d}$&  $ {\cal O}_T^{1^-},\eta_5 $&$46\to 60$&$0.14\to 0.18$&841(117)&$0.15\to 0.18$&6212(172)&&&--\\
  $T_{cc\bar d\bar d}$ &  $\eta_2 $&$48\to 60$&$0.13\to 0.16$&1165(107)&$0.14\to 0.16$&6362(140)&&&--\\
   $T_{cc\bar u\bar s}$& $ {\cal O}_{T^{1^-}_{us}},\eta_5$&$44\to 58$&$0.16\to 0.20$&558(76)&$0.17\to 0.20$&6064(167)&&&--\\
 $T_{cc\bar s\bar s}$&  $ \eta_2 $&$46\to 58$&$0.15\to 0.20$&1085(109)&$0.17\to 0.20$&6110(154)&&&-- \\
 
  \bf Class L&&&&&&&\\
  { \bma  $0^-$} &&&&&&&\\
$T_{cc\bar u\bar d}$&$ \eta_5$&$24\to 36$&$0.32\to 0.45$&289(42)&$0.41\to 0.47$&4380(128)&23&$0.27\to 0.40$&4410(140) \\
$T_{cc\bar d\bar d}$& $\eta_1$&$24\to 36$&$0.25\to 0.35$&402(45)&$0.25\to 0.35$&3965(76)&24&$0.26\to 0.33$&4430(120) \\
 $T_{cc\bar u\bar s}$& $ \eta_5$&$24\to 36$&$0.50\to 0.56$&236(51)&$0.50\to 0.56$& 4251(127)&24&$0.24\to 0.39$&4500(160) \\

  $T_{cc\bar s\bar s}$&$\eta_1$&$24\to 36$&$0.25\to 0.36$&322(42)&$0.25\to 0.36$&4096(96)&25&$0.25\to 0.36$&4460(130)\\
{\bma  $1^-$} &&&&&&&\\
  $T_{cc\bar u\bar d}$&  $\eta_6 $&$26\to 40$&$0.22\to 0.31$&293(51)&$0.22\to 0.31$&4113(108)&23&$0.27\to 0.32$&4340(160)\\
 $T_{cc\bar d\bar d}$& $ \eta_1 $&$26\to 40$&$0.23\to 0.32$&380(39)&$0.23\to 0.32$&4170(110)&23&$0.28\to 0.33$&4350(140) \\
   $T_{cc\bar u\bar s}$& $ \eta_6$&$26\to 40$&$0.23\to 0.32$&262(29)&$0.23\to 0.32$&4157(100) &23&$0.27\to 0.35$&4350(160)\\

 $T_{cc\bar s\bar s}$&  $ \eta_1 $&$26\to 40$&$0.26\to 0.35$&289(37)&$0.26\to 0.35$&4210(103)&24&$0.27\to 0.36$&4470(130) \\

   \hline\hline
\end{tabular}}
\label{tab:restccmc}
\end{center} 
\end{table*} 
\begin{table*}[hbt]
\begin{center}
\setlength{\tabcolsep}{0.7pc}
\caption{ \scriptsize $T_{bb\bar{q}\bar{q}'}$ predictions of the couplings [keV] and masses [MeV] compared with the ones in Ref.\,\cite{DCCZ}.}
\vspace*{-0.25cm} 
 {\scriptsize
  \begin{tabular}{lllllll | lll}
&\\
\hline
\hline
\multicolumn{7}{c}{\bf Our Work}&\multicolumn{3}{c}{Ref.\,\cite{DCCZ}}\\
\hline
States&Current  &$t_c$ [GeV$^2$] &$\tau$  [GeV$^{-2}$]  & $f^{NLO} _{T_{bbqq'}}$&$\tau$  [GeV$^{-2}$] &$M^{NLO}_{T_{bbqq'}}$&$t_c$ & $\tau$  [GeV$^{-2}$] &$M^{LO}_{T_{ccqq'}}$\\ 
 \hline 
 \bf Class H &&&&&&&\\
 { \bma $0^-$} &&&&&&&\\
$T_{bb\bar u\bar d}$&$ {\cal O}_{T_{ud}^{0^-}},\eta_4$&$185\to 230$&$0.060\to 0.075$&424(108)&$0.060\to 0.075$&13100(278)&--&&  ?\\
$T_{bb\bar u\bar d}$&$ \eta_5$&$180\to 220$&$0.080\to 0.090$&165(31)&$0.080\to 0.095$&12730(207)&115&$0.13 \to 0.14$& 10300(300) \\
$T_{bb\bar d\bar d}$ & $\eta_2$&$185\to 230$&$0.060\to 0.080$&373(90)&$0.065\to 0.080$&13039(258)&&&--- \\
 $T_{bb\bar u\bar s}$& $ {\cal O}_{T^{0^-}_{us}},\eta_4$&$185\to 230$&$0.070\to 0.080$&283(65)&$0.070\to 0.080$&13005(247)&&&-- \\
  $T_{bb\bar u\bar s}$& $ \eta_5$&$175\to 210$&$0.085\to 0.090$&144(40)&$0.080\to 0.100$& 12553(252)&115&$0.13 \to 0.14$&10400(200) \\
  $T_{bb\bar s\bar s}$ & $\eta_2$&$185\to 230$&$0.060\to 0.085$&302(82)&$0.080\to 0.085$&12745(274)&&&--- \\
    {\bma $1^-$} &&&&&&&\\
$T_{bb\bar u\bar d}$&$ {\cal O}_{T_{ud}^{1^-}},\eta_5$&$190\to 230$&$0.065\to 0.080$&246(59)&$0.065\to 0.080$&13053(262)&&&?\\
$T_{bb\bar d\bar d}$ & $\eta_2$&$190\to 230$&$0.065\to 0.075$&361(75)&$0.065\to 0.075$&13131(240)&&&-- \\
 $T_{bb\bar u\bar s}$& $ {\cal O}_{T^{1^-}_{us}},\eta_5$&$180\to 230$&$0.070\to 0.085$&158(41)&$0.070\to 0.085$&12910(268)&&&-- \\

  $T_{bb\bar s\bar s}$ & $\eta_2$&$180\to 230$&$0.075\to 0.090$&294(69)&$0.075\to 0.090$&12834(246)&&&-- \\

  \bf Class L&&&&&&&\\
  { \bma $0^-$} &&&&&&&\\
$T_{bb\bar d\bar d}$& $\eta_1$&$130\to 160$&$0.120\to 0.160$&14(2)&$0.120\to 0.160$&10407(124)&125&$0.10 \to 0.14$& 10600(300) \\
$T_{bb\bar s\bar s}$&$\eta_1$&$130\to 160$&$0.130\to 0.170$&12(2)&$0.130\to 0.170$&10437(109)&125&$0.10 \to 0.15$& 10600(300)\\
  { \bma $1^-$} &&&&&&&\\
  $T_{bb\bar u\bar d}$&  $\eta_6 $&$130\to 160$&$0.110\to 0.150$&11(2)&$0.110\to 0.150$&10469(110)&120&$0.11 \to 0.14$& 10400(200)\\
 $T_{bb\bar d\bar d}$& $ \eta_1 $&$130\to 160$&$0.120\to 0.160$&12(2)&$0.120\to 0.160$&10412(99)&125&$0.11 \to 0.14$& 10600(300) \\
   $T_{bb\bar u\bar s}$& $ \eta_6$&$130\to 160$&$0.120\to 0.160$&10(2)&$0.120\to 0.160$&10438(97)&120&$0.11 \to 0.14$&10400(200)\\

 $T_{bb\bar s\bar s}$&  $ \eta_1 $&$130\to 160$&$0.130\to 0.170$&10(2)&$0.130\to 0.170$&10445(116)& 125 &$0.10 \to 0.15$& 10600(300) \\

   \hline\hline
\end{tabular}}
\label{tab:restccmb}
\end{center}
\end{table*}

\end{document}